\documentstyle{article}
\date{}
\def\be{\begin{equation}}
\def\ee{\end{equation}}
\def\bea{\begin{eqnarray}}
\def\eea{\end{eqnarray}}
\def\s{\sigma}
\def\al{\alpha}
\def\la{\lambda}
\def\de{\delta}
\def\om{\omega}
\title{Rotational stability of linear string baryon configuration}
\author{ V.P.\,Petrov, G.S.\,Sharov\thanks{E-mail: german.sharov@tversu.ru}\\
{\small Tver state university}\\
{\small Tver, 170002, Sadovyj per. 35, Mathem. dep-t.}}
\begin{document}
\maketitle
\begin{abstract}
For the linear baryon string model with three massive points
(three quarks) connected sequentially by the relativistic strings
the initial-boun\-dary value problem is stated and solved in general.
This problem implies defining a classical motion of the system on the base
of given initial position and initial velocities of string points.
The given solution work reduces the initial-boundary value problem
for the considered model to the system of ordinary differential
equations that can be integrated numerically in general.
The examples of numerical simulation of the system motion are considered,
which, in particular, allow to answer to the old question about stability
of a flat uniform rotation of the linear string configuration.
\end{abstract}

\bigskip
\noindent{\bf Introduction}
\medskip

String models of baryon differ from each other by geometric
character of junction of three massive points by relativistic strings.
Four variants are possible: a) the ``three-string" model
or Y-configuration with three strings from three quarks joined
in the fourth massless point \cite{AY,PY}; b) the ``triangle" model
or $\Delta$-configuration with pairwise connection of three quarks
by three relativistic strings \cite{Tr,PRTr}; c) the quark-diquark
model q-qq \cite{Ko} (from the point of view of classical dynamics it
coincides with the meson model of relativistic string with massive
ends \cite{Ch,BN}); d) the linear configuration q-q-q with
quarks connected in series \cite{4B}.

In the present work the latter model is considered.
It is not investigated practically in comparison with others.
The common opinion was that the configuration q-q-q is unstable.
It seemed obvious, that for the rotational motion of this system
the centrifugal force rejects the middle quark at an end,
and the system q-q-q transforms in quark-diquark q-qq \cite{Ko}.
However this conclusion remained only speculative till now ---
for its proof it was necessary to develop any way of solution
of the initial-boundary value problem for the given model
with arbitrary initial conditions. It would allow us to investigate
on stability, for example, the known rotational motion, in which
the rectilinear string uniformly rotates and the middle quark is
at rest at the center of rotation.

In this work the initial-boundary value problem for classical
motion of the linear configuration q-q-q is solved on the base
of approach, applied earlier to the ``meson" model of relativistic string
with massive ends \cite{BSh}.

The initial-boundary value problem for the baryonic model has
additional difficulties in comparison with the mesonic one.
In particular, for an arbitrary motion of the q-q-q system
the world surface of the string is not smooth on the trajectory
the middle quark, moreover, there is no parametrization with the
conformally flat induced metric and with the quark trajectories as
coordinate lines.

These problems and also equations of motion and boundary conditions
for the given model are considered in Sect. 1.
For solving this initial-boundary value problem
the Fourier method (widely used in the string theory \cite{GSW})
is not applicable owing to nonlinearity of the boundary conditions
and necessity of realization of the orthonormality conditions. Similar difficulties connected with the latter conditions arise in numerical
solution of the problem with the help of difference schemes.
In Sect. 2 we suggest another approach to solving the given problem.
In section 3 with the help of numerical methods the concrete
example of motion for the q-q-q string baryon model is investigated.
Driving for.

\bigskip
\noindent{\bf 1. Equations of motion and boundary conditions\newline
On quark trajectories}
\medskip

Let's consider an open relativistic string with the tension $\gamma$
carrying three pointlike masses $m_1$, $m_2$, $m_3$ (the masses $m_1$
and $m_3$ are plased at the ends of the string).
The action for this system is
\be S[X^\mu]=-\int\limits_{\tau_1}^{\tau_2}\! d\tau\left\{\gamma\!
\int\limits_{\s_1(\tau)}^{\s_3(\tau)}\!\!
\left[\langle\dot X,X'\rangle^2-\dot X^2X'{}^2\right]^{\frac12}\!d\s+\sum
_{i=1}^3m_i\sqrt{X_i^{*2}(\tau)}\right\}.\ee
Here $X^\mu(\tau,\s)$ are coordinates of a point of
the string in $D$-dimensional Minkowski space $R^{1,D-1}$,
the speed of light $c=1$,
$\,(\tau,\s)\in\Omega=\Omega_1\cup\Omega_2$ (Fig.\,1),
$\langle a,b\rangle=a^\mu b_\mu$ --- (pseudo)scalar product,
$\dot X^\mu=\partial_\tau X^\mu$, $X'{}^\mu=\partial_\s X^\mu$,
$X_i^{*\mu}(\tau)=\frac d{d\tau}X^\mu(\tau,\s_i(\tau))$; $\s_i(\tau)$
($i=1,2,3$) --- inner coordinates of world lines of pointlike masses
(quarks).

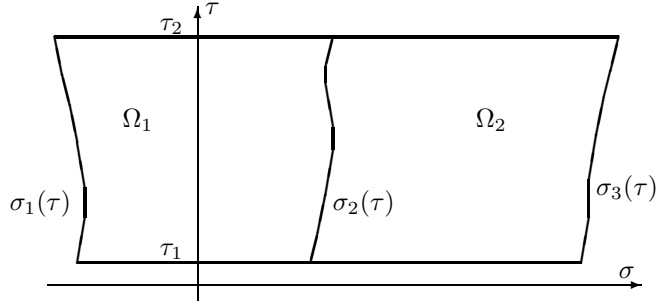
\begin{figure}[th]
\begin{center}
\begin{picture}(86,40)
\put(5,2){\vector(1,0){79}} \put(25,0){\vector(0,1){39}}
\thicklines
\put(9,5){\line(1,0){67}} \put(6,35){\line(1,0){75}}
\put(9,5){\line(1,6){1}} \put(10,11){\line(0,1){4}}
\put(10,15){\line(-1,6){1}} \put(9,21){\line(-1,5){1}}
\put(8,26){\line(-1,4){1}} \put(7,30){\line(-1,5){1}}
\put(40,5){\line(1,4){1}} \put(41,9){\line(1,5){1}}
\put(42,14){\line(1,6){1}} \put(43,20){\line(0,1){3}}
\put(43,23){\line(-1,6){1}} \put(42,29){\line(0,1){2}}
\put(42,31){\line(1,4){1}}
\put(76,5){\line(1,6){1}} \put(77,11){\line(0,1){5}}
\put(77,16){\line(1,6){1}} \put(78,22){\line(1,5){1}}
\put(79,27){\line(1,4){2}}
\put(81,3){$\sigma$} \put(26,38){$\tau$}
\put(20,36){$\tau_2$} \put(20,6){$\tau_1$}
\put(0,12){$\sigma_1(\tau)$} \put(43,12){$\sigma_2(\tau)$}
\put(78,14){$\sigma_3(\tau)$} \put(15,23){$\Omega_1$}
\put(62,23){$\Omega_2$}
\end{picture}
\caption{Domain of integration in (1).}
\end{center}
\end{figure}

We use the notations
$$\;L(\dot X,X')=-\gamma\sqrt{\langle\dot X,X'\rangle^2-
\dot X^2X'{}^2},\quad\Lambda_i(X_i^*)=-m_i\sqrt{X_i^*{}^2}.$$
Varying action (1) and equating the variation
$\de S[X^\mu]$ to zero we shall receive the equations of motion
\be
\frac\partial{\partial\tau}\frac{\partial L}{\partial\dot X^\mu}
+\frac\partial{\partial\s}\frac{\partial L}{\partial X'{}^\mu}=0,
\qquad(\tau,\s)\in\Omega.\ee
$\s=\s_i(\tau)$.
and boundary conditions at the quark trajectories $\s=\s_i(\tau)$.

To derive boundary conditions in the model ``q-q-q" we are to take into
account the discontinuities of $\dot X^\mu$, $X'{}^\mu$ on the line
$\sigma=\sigma_2(\tau)$. Thereby the term
\be\int\!\!\!\int_\Delta\bigl[\frac\partial{\partial\tau}\bigl
(\frac{\partial L}{\partial\dot X^\mu}\delta X^\mu\bigr)+
\frac\partial{\partial\sigma}\bigl(\frac{\partial L}
{\partial X'{}^\mu}\delta X^\mu\bigr)\bigr]\,d\tau\,d\sigma
\ee
in $\delta S[X^\mu]$ transformed using the Green's formula equals
the sum of two curvilinear integrals of internal boundary values
along the borders of the domains $\Omega_1$, $\Omega_2$ (Fig.1)
and therefore --- in the following boundary conditions:
\bea
&&\frac d{d\tau}\frac{\partial\Lambda_1}{\partial X_1^{*\mu}}-
\biggl[\frac{\partial L}{\partial X'{}^\mu}-\frac{\partial L}
{\partial\dot X^\mu}\s_1'(\tau)\biggr]\bigg|_{\s=\s_1(\tau)}=0,\nonumber\\
&&\begin{array}{c}\displaystyle
\frac d{d\tau}\frac{\partial\Lambda_2}{\partial X_2^{*\mu}}-
\biggl[\frac{\partial L}{\partial X'{}^\mu}-\frac{\partial L}
{\partial\dot X^\mu}\s_2'(\tau)\biggr]\bigg|_{\s=\s_2(\tau)+0}+{}\\
\displaystyle\quad{}+\biggl[\frac{\partial L}{\partial X'{}^\mu}-
\frac{\partial L}{\partial\dot X^\mu}\s_2'(\tau)\biggr]
\bigg|_{\s=\s_2(\tau)-0}=0,\end{array}\\
&&\frac d{d\tau}\frac{\partial\Lambda_3}{\partial X_3^{*\mu}}+
\biggl[\frac{\partial L}{\partial X'{}^\mu}-\frac{\partial L}
{\partial\dot X^\mu}\s_3'(\tau)\biggr]\bigg|_{\s=\s_3(\tau)}=0,\nonumber
\eea

The equations of motion of the string and the boundary conditions have
the simplest form if with the help of nondegenerate reparametrization
$\tau=\tau(\tilde\tau,\tilde\s)$, $\s=\s(\tilde\tau,\tilde\s)$
the induced metric on the world surface of the string is made
continuous and conformally-flat \cite{Tr},i.e., satisfies the orthonormality
conditions.
\be\dot X^2+X'{}^2=0,\qquad\langle\dot X,X'\rangle = 0.\ee

Under conditions (5) the equations of motion (2) become linear
\be \ddot X^\mu-X''{}^\mu=0\ee
and the boundary conditions (4) take the simplest form
\bea
&& m_1\frac d{d\tau}U^\mu_1(\tau)
-\gamma\bigl(X'{}^\mu+\s_1'(\tau)\,\dot X^\mu\bigr)\bigg|_{\s=\s_1(\tau)}=0,
\nonumber\\
&&\begin{array}{c}\displaystyle m_2\frac d{d\tau}U^\mu_2(\tau)+
\gamma\big(X'{}^\mu+\s_2'(\tau)\,\dot X^\mu\big)\bigg|_{\s=\s_2(\tau)-0}-{}\\
\displaystyle\quad{}-\gamma\big(X'{}^\mu+\s_2'(\tau)\,\dot X^\mu\big)
\bigg|_{\s=\s_2(\tau)+0}=0,\end{array}\\
&& m_3\frac d{d\tau}U^\mu_3(\tau)+\gamma\bigl(X'{}^\mu+\s'_3(\tau)\,
\dot X^\mu\bigr)\bigg|_{\s=\s_3(\tau)}=0.\nonumber
\eea
Here the notations
\be
U^\mu_i(\tau)=\frac{X_i^{*\mu}(\tau)}{\sqrt{X_i^*{}^2(\tau)}}=
\frac{\dot X^\mu+\s_i'(\tau)\,X'{}^\mu}
{\sqrt{\dot X^2\cdot(1-\s_i'{}^2)}}\bigg|_{\s=\s_i(\tau)},\quad i=1,\,2,\,3
\ee
for unit $R^{1,D-1}$-velocity vector of i-th quark was taken.

From the physical point of view equations (4) are the 2-nd
Newtonian law for the material points $m_i$.

Eqs.~(5) and (6) are invariant with respect to reparametrizations
$\tau\pm\sigma=f_\pm(\tilde\tau\pm\tilde\sigma)$ \cite{BN}. Choosing
these two arbitrary functions $f_\pm$ one can fix two (of three)
functions $\sigma_i(\tau)$, for example, in the form
\begin{equation}
\sigma_1(\tau)=0,\qquad\sigma_3(\tau)=\pi.\end{equation}
The first equation (9) may be obtained at the first step
by using the above reparametrization with required $f_+$
and $f_-(\eta)= \eta$, the second equation (9) ---
at the second step with $f_+=f_-$.
This procedure doesn't permit to fix $\sigma_2={}$const
for all $\tau$ in general.

For the initial-boundary value problem the function
$\sigma_2(\tau)$ should be is calculated on the base
of initial data.

The function $X^\mu(\tau,\s)$
is continuous in $\Omega$ but on the line $\s_2(\tau)$ its derivatives
can have discontinuities of 1-st type (except for tangential $V_i^\mu$ and
$\frac d{d\tau}V_i^\mu$).

\bigskip
\noindent{\bf 2. Initial-boundary value problem }
\medskip

The initial-boundary value problem for a relativistic string with the action
(1) we shall formulate in general as follows: `` to find a solution
of the equation (6) $X^\mu(\tau,\s)$, $\mu=0,1,\dots,D-1$, smooth in the
domain
\be \Omega={(\tau,\s):\,\tau>\tau_0(\s);\;\;
\s_1(\tau)<\s<\s_3(\tau)}.\ee
continuously differentiable on $\partial\Omega $ satisfying
to the orthonormality conditions (5) and boundary conditions (7) and
two given initial conditions: an initial position of the string
in Minkowski space and an initial velocity of a point of the string".

An initial position of the string can be given as a
spacelike curve in Minkowski space
\be x^\mu=\rho^\mu(\la), \quad\la\in [\la_1,\la_3],\ee
satisfying the conditions
$[\rho^\prime(\la)]^2<0$. Initial velocities on the
initial curve can be given by timelike
$R^{1,D-1}$\,-\,vector $\,v^\mu(\la)$, $\la\in[\la_1,\la_3]$.

To solve the problem we set the initial curve (11)
on the world surface of the string parametrically
$$\tau=\tau(\la), \quad\s=\s(\la), \qquad\la\in[\la_1,\la_3]$$
(in expression (10) the same line is given as $\tau=\tau_0(\s)$).

In these notations the initial conditions (initial position of a string and
the initial velocities) can be given in general \cite{BSh,JVM}
\bea
&X^\mu\bigl(\tau(\la),\s(\la)\bigr)=\rho^\mu(\la),\qquad
\la_1\leq\la\leq\la_3,&\\
&\!\al(\la)\,\dot X^\mu\bigl(\tau(\la),\s(\la)\bigr)+\beta(\la)\,
X'^\mu\bigl(\tau(\la),\s(\la)\bigr)=\rho^\mu(\la),\quad
\la_1\leq\la\leq\la_3,\;\;&\eea
where $\al(\la)$, $\beta(\la)$ are arbitrary functions
satisfying the inequality $ \alpha (\la) > | \beta (\la) | $
and (if the relations (9) are tied) the condition on at the ends
$\beta(\la_1)=\beta(\la_3)=0$.

Functions $\tau(\la)$, $\s(\la)$, $\al(\la)$, $\beta(\la)$ are related
by formulas \cite{BSh}

\be \s^\prime(\la)=\frac{\alpha\Delta+\beta P}{v^2},\qquad
\tau^\prime(\la)=\frac{\beta\Delta+\alpha P}{v^2},\ee
where $\,P(\la)=\left\langle v(\la),\rho^\prime(\la)\right\rangle,\quad
\Delta(\la)=\sqrt{{\langle v,\rho^\prime\rangle}^2-v^2\rho^{\prime2}}$.
It is convenient to choose the constants of integration in (14) so
that $\tau(\la_1)=\s(\la_1)=0$ (Fig.\,2).
Realization of the condition $\s(\la_3)=\pi$ is possible
by multiplying $\alpha(\la)$ and $\beta(\la)$ by a constant.

\begin{figure}[ht]
\begin{center}
\begin{picture}(110,60)
\put(5,2){\vector(1,0){100}} \put(10,0){\vector(0,1){59}}
\put(8,3){0} \put(105,3){$\sigma$} \put(11,59){$\tau$}
\put(-1,20){$\sigma_1=0$} \put(55,29){$\sigma_2(\tau)$}
\put(101,20){$\sigma_3=\pi$} \put(31,37){$\Omega_1$}
\put(75,37){$\Omega_2$} \put(31,15){$A$} \put(75,17){$B$}
\put(20,3){$\s=\s(\la)$} \put(20,6){$\tau=\tau(\la)$}
\put(34,4){\line(6,-1){6}}
\put(10,2){\line(1,1){45}} \put(53,5){\line(1,1){47}}
\put(53,5){\line(-1,1){43}} \put(100,4){\line(-1,1){45}}
\thicklines
\put(10,2){\line(0,1){56}} \put(100,4){\line(0,1){54}}
\put(10,2){\line(5,-1){5}} \put(15,1){\line(6,-1){6}}
\put(21,0){\line(1,0){4}} \put(25,0){\line(6,1){6}}
\put(31,1){\line(5,1){10}} \put(41,3){\line(6,1){24}}
\put(65,7){\line(1,0){18}} \put(83,7){\line(6,-1){12}}
\put(95,5){\line(5,-1){5}}
\put(53,5){\line(-1,6){1}} \put(52,11){\line(0,1){6}}
\put(52,17){\line(1,6){2}} \put(54,29){\line(1,5){1}}
\put(55,34){\line(1,6){1}} \put(56,40){\line(0,1){3}}
\put(56,43){\line(-1,6){2.5}}
\end{picture}
\caption{Начальная кривая в $\Omega$.}
\end{center}\end{figure}

The line $\s_2(\tau)$ devides domains $\Omega$ into two domains
$\Omega_1$ and $\Omega_2$.
Let's consider the common solutions of the equation (6) in these domains
\be\begin{array}{ll}
X^\mu(\tau,\s)=\frac1{2}\bigl[\Psi^\mu_{1+}(\tau+\s)+\Psi^\mu_{1-}(\tau-\s)
\bigr],& (\tau,\s)\in\Omega_1\\
X^\mu(\tau,\s)=\frac1{2}\bigl[\Psi^\mu_{2+}(\tau+\s)+\Psi^\mu_{2-}(\tau-\s)
\bigr],& (\tau,\s)\in\Omega_2.\rule[3mm]{0mm}{1mm}\end{array}\ee

We can find the functions
 $\Psi^{\prime\mu}_{i\pm}$, $i=1,2$ using the formulas \cite{BSh}
\be \Psi^{\prime\mu}_{i\pm}\bigl(\tau(\la)\pm\s(\la)\bigr)=
\frac{(\Delta\mp P)v^\mu\pm v^2\rho^{\prime\mu}}
{\Delta\bigl(\alpha(\la)\pm\beta(\la)\bigr)}\ee
If (as mentioned above) we choose
$$\tau(\la_1)=\s(\la_1)=0,\qquad\s(\la_3)=\pi,$$
that the formulas (16) will determine
 $\Psi^\mu_{i\pm}(\xi)$ in the following segments:
\be\!\!\!\!\!\!\begin{array}{ll}
\Psi^\mu_{1+}(\xi),\;\xi\in\bigl[0,\tau(\la_2)+\s(\la_2)\bigr],&
\Psi^\mu_{2+}(\xi),\;\xi\in\bigl[\tau(\la_2)+\s(\la_2),
\tau(\la_3)+\pi\bigr],\\
\Psi^\mu_{1-}(\xi),\;\xi\in\bigl[\tau(\la_2)-\s(\la_2),0\bigr],&
\Psi^\mu_{2-}(\xi),\;\xi\in\bigl[\tau(\la_3)-\pi,\tau(\la_2)-
\s(\la_2)\bigr],\rule[3mm]{0mm}{1mm}\end{array}\!\!\ee
that allows us to find the solution of the problem as (15) in the zones
$A=\bigl\{(\tau,\s):\tau(\la(\s))<\tau\leq\s;\tau+\s\leq\tau(\la_2)
+\s(\la_2)\bigr\}$ и
$B=\bigl\{(\tau,\s):\tau(\la(\s))<\tau\leq\s+\tau(\la_2)-
\s(\la_2);\tau+\s\leq\tau(\la_3)+\pi\bigr\}$ (Fig.\,2).
In these zones the solution depends only on initial data
without influence of the boundaries.
The constant of the integration in Eq. (16) is
determined with the help of the initial condition (12).

In others zones of the domains $\Omega_1$ and $\Omega_2$ we shall
Obtain the solution by prolongating the functions
$\Psi^\mu_{i\pm}$ on all positive semiaxis with the help
of the boundary conditions.

For the given initial-boundary value problem the boundary conditions
look like (7). By choosing the inner equations of the string ends
as (9) and by substituting in Eqs. (7) common solutions (15)
we transform this system (it is supposed
that the masses $m_1$, $m_2$, $m_3$ are final and nonzero):
\bea
&\displaystyle\frac{dU^\mu_1(\tau)}{d\tau}-\frac{\gamma}{2m_1}
\bigl[\Psi^{\prime\mu}_{1+}(\tau)-\Psi^{\prime\mu}_{1-}(\tau)\bigr]=0,&\\
&\begin{array}{c}\displaystyle
\frac{dU^\mu_2(\tau)}{d\tau}+\frac{\gamma}{2m_2}\bigg[(1+\s'_2)\big[
\Psi^{\prime\mu}_{1+}(\tau+\s_2)-\Psi^{\prime\mu}_{2+}(\tau+\s_2)\big]+{}\\
\displaystyle{}+(1-\s'_2)\big[\Psi^{\prime\mu}_{2-}(\tau-\s_2)-
\Psi^{\prime\mu}_{1-}(\tau-\s_2)\big]\bigg]=0,\end{array}&\\
&\displaystyle\frac{dU^\mu_3(\tau)}{d\tau}+\frac{\gamma}{2m_3}\bigl[
\Psi^{\prime\mu}_{2+}(\tau+\pi)-\Psi^{\prime\mu}_{2-}(\tau-\pi)\bigr]=0.&
\eea
The expressions for the unit velocity vectors (8) after substitutions (9),
(16) may be rewritten as
\bea
\!U^\mu_1(\tau)=\frac{\Psi^{\prime\mu}_{1+}(\tau)+
\Psi^{\prime\mu}_{1-}(\tau)}{\sqrt{2\bigl\langle\Psi^\prime_{1+}(\tau),
\Psi^\prime_{1-}(\tau)\bigr\rangle}},\;\;
U^\mu_3(\tau)=\frac{\Psi^{\prime\mu}_{2+}(\tau+\pi)+\Psi^{\prime\mu}_{2-}
(\tau-\pi)}{\sqrt{2\bigl\langle\Psi^\prime_{2+}(\tau+\pi),\Psi^\prime_{2-}
(\tau-\pi)\bigr\rangle}},&&\nonumber\\
\begin{array}{c}\displaystyle
U^\mu_2(\tau)=\frac{(1+\s'_2)\,\Psi^{\prime\mu}_{1+}(\tau+\s_2)+(1-\s'_2)\,
\Psi^{\prime\mu}_{1-}(\tau-\s_2)}{\sqrt{2(1-\s^{\prime2}_2)\big\langle
\Psi^\prime_{1+}(\tau+\s_2),\Psi^\prime_{1-}(\tau-\s_2)\big\rangle}}={}\\
\displaystyle{}=\frac{(1+\s'_2)\,\Psi^{\prime\mu}_{2+}(\tau+\s_2)+
(1-\s'_2)\,\Psi^{\prime\mu}_{2-}(\tau-\s_2)}{\sqrt{2(1-\s^{\prime2}_2)
\big\langle\Psi^\prime_{2+}(\tau+\s_2),\Psi^\prime_{2-}(\tau-\s_2)\big\rangle}}.
\end{array}\qquad\qquad&&
\eea
In last expression the continuity of the function $X^\mu(\tau,\s)$
and its tangential derivatives on the line $\s=\s_2(\tau)$
was taken into account. By differentiating on $\tau$ the identity
$X^\mu\big|_{\s=\s_2(\tau)-0}=X^\mu\big|_{\s=\s_2(\tau)+0}$
(after substition (15)) we obtain
\be(1+\s'_2)\,\Psi^{\prime\mu}_{1+}(+)+(1-\s'_2)\,\Psi^{\prime\mu}_{1-}(-)=
(1+\s'_2)\,\Psi^{\prime\mu}_{2+}(+)+(1-\s'_2)\,\Psi^{\prime\mu}_{2-}(-).
\ee
Here the notations
$(+)\equiv(\tau+\s_2(\tau))$, $(-)\equiv(\tau-\s_2(\tau))$.

We transform the systems of the ordinary differential equations
(18)\,-\,(21) to the normal form by the method used in Ref \cite{BSh}.
Let's consider transformation of the boundary condition (19)
with the help given method.

Using the scalar product of the Eq. (21) by
$(1+\s'_2)\,\Psi^{\prime\mu}_{i+}(+)$ and
$(1-\s'_2)\,\Psi^{\prime\mu}_{i-}(-)$ and taking into account
the isotropy of the vectors $\Psi^{\prime\mu}_{i\pm}$
$$\Psi^{\prime2}_{i+}(\tau)=\Psi^{\prime2}_{i-}(\tau)=0$$
[resulting from conditions (5)] we obtain the relations
\be\begin{array}{c}
\frac1{\sqrt2}{\sqrt{(1-\s^{\prime2}_2)\big\langle\Psi^\prime_{1+}(+),
\Psi^\prime_{1-}(-)\big\rangle}}
=(1+\s'_2) \big\langle U_2(\tau),\Psi'_{1+}(+)\big\rangle=\\
=(1-\s'_2)\big\langle U_2(\tau),\Psi'_{1-}(-)\big\rangle=
\frac1{\sqrt2}{\sqrt{(1-\s^{\prime2}_2)
\big\langle\Psi^\prime_{2+}(+),\Psi^\prime_{2-}(-)\big\rangle}}=\\
=(1+\s'_2)\big\langle U_2(\tau),\Psi'_{2+}(+)\big\rangle=
(1-\s'_2)\big\langle U_2(\tau),\Psi'_{2-}(-)\big\rangle.\rule[3mm]{0mm}{1mm}
\end{array}\ee

The equalities (21) and (23) result in the relations
$$(1+\s'_2)\,\Psi^{\prime\mu}_{i+}(+)+(1-\s'_2)\,\Psi^{\prime\mu}_{i-}(-)=
2(1\pm\s'_2)\big\langle U_2(\tau),\Psi^\prime_{j\pm}(\pm)\big\rangle U_2^\mu(\tau),$$
(right for all 8 variants of a choice of indexes $i$, $j$ and signs $\pm$)
with the help of which we shall transform the summands in  Eq. (19) as follows:
$$\begin{array}{c}
(1+\s'_2)\,\Psi^{\prime\mu}_{1+}(+)-(1-\s'_2)\,\Psi^{\prime\mu}_{1-}(-)=
2(1-\s'_2)\big[\langle U_2,\Psi^\prime_{1-}(-)\rangle U_2^\mu-
\Psi^{\prime\mu}_{1-}(-)\big],\\
(1-\s'_2)\,\Psi^{\prime\mu}_{2-}(-)-(1+\s'_2)\,\Psi^{\prime\mu}_{2+}(+)=
2(1+\s'_2)\big[\langle U_2,\Psi^\prime_{2+}(+)\rangle U_2^\mu-
\Psi^{\prime\mu}_{2+}(+)\big].\rule[3mm]{0mm}{1mm}
\end{array}$$

The equation (19) take the normal form now
\be\frac{dU^\mu_2}{d\tau}=\frac{\gamma}{m_2}\big[\de^\mu_\nu-U^\mu_2(\tau)\,
U_{2\nu}(\tau)\big]\big[(1+\s'_2)\,\Psi^{\prime\nu}_{2+}(+)+
(1-\s'_2)\,\Psi^{\prime\nu}_{1-}(-)\big],\ee
where $\de^\mu_\nu=\left\{\begin{array}{ll}1,&\mu=\nu\\0,&\mu\ne\nu.
\end{array}\right.$

Using the same receptions we shall transform the equations (18) and (20)
to the form
\bea
&\displaystyle\frac{dU^\mu_1(\tau)}{d\tau}=\frac{\gamma}{m_1}\big[\de^\mu_\nu
-U^\mu_1(\tau)\,U_{1\nu}(\tau)\big]\Psi^{\prime\nu}_{1+}(\tau),&\\
&\displaystyle\frac{dU^\mu_3(\tau)}{d\tau}=\frac{\gamma}{m_3}\big[\de^\mu_\nu
-U^\mu_3(\tau)\,U_{3\nu}(\tau)\big]\Psi^{\prime\nu}_{2-}(\tau-\pi).&
\eea

The systems (24)\,-\,(26) need the initial conditions
\be  U^\mu_i\big(\tau(\la_i)\big)=v^\mu(\la_i)\big/\sqrt{v^2(\la_i)},
\qquad i=1,2,3.\ee

Integrating systems (24)\,-\,(26) with the initial conditions (27) we can
determine for $\tau>\tau(\la_i)$ unknown vector functions
$U^\mu_i(\tau)$ with the help of the functions $\Psi^{\prime\mu}_{i\pm}$
known from the initial data on the segments (17).
This procedure is limited on $\tau$ to ordinates of points in which
the trajectories $\s=\s_i(\tau)$, $i= 1,2,3$ cross the characteristic
lines $\tau\pm\s={}$ const (Fig.\,2). However we can continue this
procedure for all $\tau$ if for every value of $\tau$ we determine
the functions $\Psi^{\prime\mu}_{i\pm}$ outside segments (17) with
the help of Eqs. (18)\,-\,(20), (24)\,-\,(26).

For this purpose we rewrite Eqs. (18) and (20) as
\bea
&&\Psi^{\prime\mu}_{1-}(\tau)=\Psi^{\prime\mu}_{1+}(\tau)-
2m_1\gamma^{-1}U^{\prime\mu}_1(\tau),\\
&&\Psi^{\prime\mu}_{2+}(\tau+\pi)=\Psi^{\prime\mu}_{2-}(\tau-\pi)-
2m_3\gamma^{-1}U^{\prime\mu}_3(\tau),
\eea
Eq. (28) corresponds to Eq. (25) and Eq. (29) --- to Eq. (26).
From Eqs. (19) and (22) we deduce two similar relations
\be\begin{array}{l}
\Psi^{\prime\mu}_{1+}\big(\tau+\s_2(\tau)\big)=
\Psi^{\prime\mu}_{2+}\big(\tau+\s_2(\tau)\big)-
m_2\big[\gamma(1+\s'_2)\big]^{-1}U^{\prime\mu}_2(\tau),\\
\Psi^{\prime\mu}_{2-}\big(\tau-\s_2(\tau)\big)=
\Psi^{\prime\mu}_{1-}\big(\tau-\s_2(\tau)\big)-
m_2\big[\gamma(1-\s'_2)\big]^{-1}U^{\prime\mu}_2(\tau).\rule[3mm]{0mm}{1mm}
\end{array}\ee

For solving the system (24) we are to determine the function $\s_2(\tau)$
for $\tau>\tau(\la_2)$. This function can be obtained
by taking two of the equalities (23)
$$(1-\s'_2)\langle U_2(\tau),\Psi'_{1-}(-)\rangle
=(1+\s'_2)\langle U_2(\tau),\Psi'_{2+}(+)\rangle$$
and expressing $\s'_2$ as follows:
\be\frac{d\s_2(\tau)}{d\tau}=
\frac{\langle U_2(\tau),[\Psi'_{1-}(\tau-\s_2)-\Psi'_{2+}(\tau+\s_2)]\rangle}
{\langle U_2(\tau),[\Psi'_{1-}(\tau-\s_2)+\Psi'_{2+}(\tau+\s_2)]\rangle}.\ee

The equations and systems (24)\,-\,(31) allow us
to continue the functions $\Psi^\mu_{i\pm}$ determined from the initial
conditions with the help of formulas (16) on the segments (17).
The constants of integration in $\Psi^{\prime\mu}_{i\pm}$
are determined from the initial condition (12).
The descibed above algorithm of calculating
$\Psi^\mu_{i\pm}$ (with taking into account Eq.~(15)) is the algorithm
of solving the considered initial-boundary value problem with arbitrary
initial conditions $\rho^\mu(\la)$, $v^\mu(\la)$.

\bigskip
\noindent{\bf 3. Numerical solution of initial-boundary value problem}
\medskip

To the present time only two elementary motions of string
configurations q-q-q are known. They are analytical solutions of
equations (6) and satisfy conditions (5), (7). One of these motions is
spatially one-dimensional compression of a symmetric
($m_1=m_3$, $m_2$ is at the center) rectilinear string that
in some initial instant was at rest \cite{BN}.
The second motion is a flat uniform rotation of the rectilinear
string with the middle quark at rest at a center of rotation
(in the reference frame of this center) \cite{Ko,4B,PSh}. If the choice
of $\tau,\,\s$ satisfies the conditions (5) and (9) then this solution
can be presented as
\be
X^\mu=\om^{-1}\bigl\{a\tau;\,\sin(a\s-\phi_0)\cdot\cos a\tau;
\,\sin(a\s-\phi_0)\cdot\sin a\tau\bigr\}.\ee
Three coordinates of the vector $X^\mu$ are
$X^0=t$, $X^1=x$, $X^2=y$.
The trajectory of the middle quark on a plane $\tau,\s$ has the form
$\s_2(\tau)=\phi_0/a={}$const and the trajectories of the others
quarks have the form (9).

The vector function (32) is the solution of the equations (6),
satisfies the conditions (5) and under the relations
\be\frac\gamma{\om m_1}=\frac{\sin\phi_0}{\cos^2\phi_0},\quad
\frac\gamma{\om m_3}=\frac{\sin(a\pi-\phi_0)}{\cos^2(a\pi-\phi_0)},\quad
\cos(a\s_i-\phi_0)>0
\ee
it satisfies the boundary conditions (7).

The solution (32) describes a rotation of the rectilinear string with
an angular frequency $\om$. The quarks move at constant speeds
\be v_i=|\sin(a\s_i-\phi_0)|,\qquad i=1,2,3\ee
along the circles with the radii $R_i=v_i/\om$.
The relations (33) may be rewritten as
\be\frac\gamma{m_i}=\frac{\om v_i}{1-v_i^2}=\frac{v_i^2}{R_i(1-v_i^2)},
\quad i=1,3.
\ee

The expression (32) can be obtained as the solution of the initial-boundary
value problem (6), (7), (12), (13) with the initial conditions
\be\begin{array}{l}
\rho^\mu(\la)=\big\{0;\,-R_1+(R_1+R_3)\,\la;\,0\big\},\\
v^\mu(\la)=\big\{1;\,0;\,-v_1+(v_1+v_3)\,\la\big\},\rule[3mm]{0mm}{1mm}
\end{array}\quad\la\in[0,1],\quad\la_2=\frac{R_1}{R_1+R_3}.
\ee
If the parameters $v_i$, $R_i$ satisfy Eqs. (35) the solution of the
initial-boundary value problem has the form (32) in all domain $\Omega$
and can be obtained with the help of algorithm described in the
previous section.

To solve the problem of stability of the solution (32) we are to consider
an initial-boundary value problem with the initial conditions $\rho^\mu(\la)$
$v^\mu(\la)$ which slightly differ from conditions (36).
In particular, it is possible to put $\la_2\ne R_1/ (R_1+R_3)=v_1/(v_1+v_3)$,
i.e. to place the middle quark not at the rotational center.

For arbitrary initial conditions the solution of the problem
becomes complicated. In particular, system (24)\,-\,(26),
(31) are solvable analytically only for exclusive cases. Therefore the
use of numerical method is necessary. It makes possible to solve
the initial-boundary value problem with any initial and boundary
conditions.

The used below procedure of solving includes three stages:

1) realization of initial data, i.e., the determinition
the functions $\Psi_{i\pm}^{\prime\mu}$ in the segments (17) by
the formulas (16) [with the help the initial conditions
$\rho^\mu(\la)$ и $v^\mu(\la)$].

2) continuating the functions
$\Psi_{i\pm}^{\prime\mu}(\xi)$ (with integrating)
with the help of boundary conditions (24)\,--\,(31);

3) searching of level lines of the world surface
 $t=X^0(\tau,\s)=\,\mbox{const}$ and drawing of their projections
on the plane $0xy$ ($x\equiv X^1$, $y\equiv X^2$).

The last item is required for the best representation
of the string motion. We show `` photographs" of its
sequential positions made at regular intervals
$\Delta t$.

The results of calculations for the q-q-q system with
$m_1=1$, $m_2=3$, $m_3=2$, $\gamma=1$ obtained with use of the MATLAB
package are shown in Fig.\,3.

\begin{figure}

\caption{ Motion of the system with initial conditions
(36), (37). $\Delta t=0.15$.}
\end{figure}

The initial conditions for the given motion have form close to (36),
where
\be\omega=2,\quad v_1\simeq0.4142,\quad v_3\simeq0.4142,\quad\la_2=0.615
\ne v_1/(v1+v3)\simeq0.637.
\ee
The values $R_1$, $v_1$ satisfy conditions (35)
however the middle quark is located not at the center of rotation.

In Figs. 3a\,-\,d on the plane $Oxy$ the positions of the system are shown
sequentially with the interval $\Delta t=0.15$ in time.
The initial position of the string (36) in Fig.\,3а is marked by number 1
and next (with growth of $t$) ``photographs" of the system
are marked by numbers 2, 4 $,\dots,$ near the third quark position.
For convenience of perception the first pointlike quark $q_1$ with mass
$m_1=1$ at the end of the string is depicted as a green square, the third
quark $q_3$ with mass $m_3=2$ --- as a red square, and positions
of the middle quark $q_2$ are marked by blue circles.

The last position of the system number 11 in Fig.\,3а is
the first one in Fig.\,3b. The system q-q-q continues rotating
as a whole in the counter-clockwise direction in Figs.\,3b\,--\,d.
At the beginning (Fig.\,3а) the motion of the system as a whole is
close to the uniform rotation (32).
Later the middle quark $q_2$ under the centrifugal force is displaced to
the quark $q_1$ but there is no junction of $q_2$ with $q_1$
and transformation of the system in the quark-diquark one \cite{Ko}.
The middle quark plays a role of rotational center for the segment
of the string $q_2$\,--\,$q_1$ (positions 26, 27 in Fig.\,3c) and then
it returns to the center of a system (Fig.\,3c, 3d). The middle quark
does not reach this center and begins again centrifugal moving away
with recurring of all cycle.

As we see, the motion is quasiperiodic. It's qualitative
features are the same for rather wide variation of parameters
of the initial-boundary value problems. In particular, decreasing
of the initial deviation of $q_2$ from the center of masses reduces in
retarding of the initial stage of motion.

If the initial conditions differ from (36) by a small nonsymmetrical correction
to the velocity $v^\mu(\la)$ (for $\la_2=\frac{R_1}{R_1+R_3}$), the motion of the system is approximately the same as in Fig.\,3.
The increase of the mass $m_2$ of the middle quark $q_2$ reduces in
Increase of $\Delta R$ --- minimal length of the segment $q_1$\,--\,$q_2$.

Energy or velocity of rotation of the system under the relations (35)
weakly influences to the character of motion.
In the non-relativistic limit $v_i\ll1$ and in
relativistic case (Fig.\,3) the period of cyclical motion of
the middle quark between the center and the end is approximately
equal to the rotational time of the string.

\bigskip
\noindent{\bf Conclusion}
\medskip

In the present work the method of solution of the initial-boundary value
problem for classical motion of the linear string baryon model q-q-q is
suggested.
This approach allows us to solve the stability problem for the known solution
(32) --- flat uniform rotation of the rectilinear string with the quark
$q_2$ at rest. This solution is unstable in Lyapunov's sense ---
the as much as small perturbations result in complicated motion
(Fig.\,3) with quasi-periodical varying of the distance between
the nearest two quarks. However the minimal value of the mentioned distance
$\Delta R$ does not equal zero, in other words, the system q-q-q is not transformed in quark-diquark (q-qq) one, as was supposed speculatively
\cite{Ko}.

The analysis of this problem does not permit us to exclude the linear
configuration q-q-q from a set of string baryon configurations (q-qq,
Y, $\Delta$) for describing baryonic orbital excitations on the Regge
trajectories. At that it is possible to use expressions for the
total energy and the angular moment of the system q-q-q,
obtained in Ref. \cite{4B} for the motion (32).
\eqnarray
E =\sum_{i=1}^3\bigg[\frac\gamma\om\arcsin v_i+
\frac{m_i}{\sqrt{1-v_i^2}}\bigg]+\Delta E_S,\nonumber \\
J=\sum_{i = 1}^3\bigg[\frac1{2\om}\bigg(\frac\gamma\om
\arcsin v_i+\frac{m_iv_i^2}{\sqrt{1-v_i^2}}\bigg)+s_i\bigg] .\nonumber
\endeqnarray
Here $\Delta E_S$ is the spin-orbital correction to the energy, $s_i$ are
projections of quark spins, $v_2=0$. These expressions are applicable
and for more complicated motions, obtained with the help of weak
perturbations of the conditions (36) (see\, Fig.\,3) because of
conservation of $E$ and $J$.
Hence, for perturbed quasiperiodic motions the graph of dependence
$J=J(E^2)$ is close to linear, that allows us to describe
baryonic Regge trajectories with the help of the model q-q-q \cite{4B}.

\end{document}